\documentclass[aip,apl,preprint,superscriptaddress,super,superbib,amsmath]{revtex4-1}
\usepackage{graphicx}

\begin{document}

\title{Enhancement of the zero phonon line emission from a single NV-center in a nanodiamond via coupling to a photonic crystal cavity}

\author{Janik Wolters} \email[Electronic mail: ]{janik@physik.hu-berlin.de}
\affiliation{Nano-Optics, Institute of Physics, Humboldt-Universit\"{a}t zu
Berlin, Hausvogteiplatz~5-7, D-10117 Berlin, Germany}

\author{Andreas W. Schell}
\affiliation{Nano-Optics, Institute of Physics, Humboldt-Universit\"{a}t zu
Berlin, Hausvogteiplatz~5-7, D-10117 Berlin, Germany}

\author{G\"{u}nter Kewes}
\affiliation{Nano-Optics, Institute of Physics, Humboldt-Universit\"{a}t zu
Berlin, Hausvogteiplatz~5-7, D-10117 Berlin, Germany}

\author{Nils N\"{u}sse}
\affiliation{Department for Micro- and Nanostructured Optical Systems,
Helmholtz-Zentrum Berlin f\"{u}r Materialien und Energie GmbH,
Albert-Einstein-Str. 15, D-12489 Berlin}

\author{Max Schoengen}
\affiliation{Department for Micro- and Nanostructured Optical Systems,
Helmholtz-Zentrum Berlin f\"{u}r Materialien und Energie GmbH,
Albert-Einstein-Str. 15, D-12489 Berlin}

\author{Henning D\"{o}scher}
\affiliation{Department Materials for Photovoltaics, Helmholtz-Zentrum Berlin
f\"{u}r Materialien und Energie GmbH, Hahn-Meitner-Platz 1, D-14109 Berlin}

\author{Thomas Hannappel}
\affiliation{Department Materials for Photovoltaics, Helmholtz-Zentrum Berlin
f\"{u}r Materialien und Energie GmbH, Hahn-Meitner-Platz 1, D-14109 Berlin}

\author{Bernd L\"{o}chel}
\affiliation{Department for Micro- and Nanostructured Optical Systems,
Helmholtz-Zentrum Berlin f\"{u}r Materialien und Energie GmbH,
Albert-Einstein-Str. 15, D-12489 Berlin}

\author{Michael Barth}
\affiliation{Nano-Optics, Institute of Physics, Humboldt-Universit\"{a}t zu
Berlin, Hausvogteiplatz~5-7, D-10117 Berlin, Germany}

\author{Oliver Benson}
\affiliation{Nano-Optics, Institute of Physics, Humboldt-Universit\"{a}t zu
Berlin, Hausvogteiplatz~5-7, D-10117 Berlin, Germany}

\begin{abstract}
Using a nanomanipulation technique a nanodiamond with a single nitrogen vacancy
center is placed directly on the surface of a gallium phosphide photonic
crystal cavity. A Purcell-enhancement of the fluorescence emission at the zero
phonon line (ZPL) by a factor of $12.1$ is observed. The ZPL coupling is a
first crucial step towards future diamond-based integrated quantum optical
devices.
\end{abstract}

\maketitle
Quantum information processing (QIP) has become one of the major research
topics in quantum optics.\cite{OBrien2010} As QIP requires the control of
complex quantum systems several research groups have been striving for ways to
realize an integrated quantum technology platform.\cite{OBrien2009} Apart from
purely semiconductor based approaches\cite{Hennessy2008} diamond has emerged as
an interesting alternative. Several optically active defect centers in diamond
have been studied on the single center level. The most promising among them for
QIP applications is the nitrogen vacancy
(NV)-center.\cite{Wrachtrup2006,DiVincenzo2010} With its triplet ground state
it is capable to encode quantum bits which can be controlled and read out by
microwave or optical fields. In ultra-pure diamond, electron spin coherence
times are in the millisecond range.\cite{Balasubramanian2009}
Apart from this, the NV-center is probably the best known solid state single
photon source operating at room temperature.\cite{Grangier2000}

However, the optical properties of the NV-centers are not ideal: The coupling
strength to the electromagnetic field is small compared to other systems like
quantum dots.\cite{Imamoglu2007} Another problem is the small Debye-Waller
factor. Even at cryogenic temperatures only a small fraction of about 3\% of
the radiation is emitted into the Fourier-limited zero phonon line (ZPL).
\cite{Santori2010}

It has been proposed to overcome these problems by coupling the NV-centers to
optical microcavities,\cite{Hollenberg2008,Santori2010} where the spontaneous
emission (SE) rate is enhanced by the Purcell effect. For a perfect spatial and
spectral matching of the emitter's dipole with the cavity, the SE rate
enhancement into the cavity mode is described by the Purcell factor
\begin{equation}
 F = \frac{3}{4\pi^2}\left(\frac{\lambda_c}{n}\right)^3\frac{Q}{V_\text{eff}},
\end{equation}
where $\lambda_c/n$ is the wavelength at the cavity resonance in the medium,
$Q$ is the cavity quality factor ($Q$-factor) and $V_\text{eff}$ is the
effective mode volume. Increasing the SE rate into the ZPL will allow the
generation of a large number of indistinguishable single photons
\cite{Hollenberg2008} needed for linear optics quantum
computation.\cite{Milburn2001}

An even stronger enhancement of the emitter-cavity coupling may bring the
system to the strong coupling regime of cavity quantum electrodynamics, which
allows the realization of quantum gates and interfaces between stationary and
flying Qbits. Several attempts have been pursued to couple NV-centers to
cavities such as microsphere resonators\cite{Wang2006,Schietinger2008},
microtoroids,\cite{Barclay2009,Gregor2009} or photonic crystal cavities
(PCC).\cite{Barth2009,Lukin2010} PCCs have the particular advantage that they
can be fabricated with high quality factors as well as small mode volumes
\cite{Akahane2005}, both being figures of merit for obtaining a large Purcell
enhancement. Furthermore, PCCs can easily be integrated into more complex
systems of coupled cavities and waveguides.\cite{Tanabe2008}

In this letter we present an experimental realization of one of the most
crucial steps for integrated diamond-based quantum technology, i.e. the
coupling of the ZPL of an NV-center in a nanodiamond to a single mode of a PCC.

To achieve this, a free-standing photonic crystal slab (lattice constant
200~nm) with a so-called L3 cavity, formed by three missing holes (see Fig.~1),
is designed using FDTD simulations (Lumerical). It is optimized to have the
fundamental mode at 640~nm, slightly above the ZPL of the NV-center. The volume
of the fundamental mode is $V_\text{eff}\approx0.75 (\lambda/n)^3$. These
structures were fabricated from a 70~nm thick heteroepitaxial gallium phosphide
(GaP) layer,\cite{Doescher2008} deposited on a Si(100) substrate by electron
beam lithography and subsequent dry-etching.

Using a home-build micro photoluminescence setup ($\mu$PL) with a numerical
aperture (NA) of $0.9$, bare cavities were analyzed utilizing the intrinsic
fluorescence light under strong excitation with a green 532~nm laser (300
$\mu$W). The intrinsic fluorescence is strongly enhanced at the cavity
resonance. All the fabricated cavities show resonances around 642~nm with a
typical deviation of $\pm2$~nm and a $Q$-factor of around 1000.

To place a single nanodiamond (height 35~nm) with a single NV-center in the
cavity, first a diamond suspension was spin-coated onto a glass cover slip.
These diamonds were then optically pre-characterized using a confocal
microscope (NA $1.35$) with a Hanbury-Brown and Twiss setup under pulsed excitation
with a 532 nm laser at a repetition rate of 10~MHz. For a diamond containing a
single NV-center the autocorrelation function $g^{(2)}(\tau)=\left\langle I(t)
I(t+\tau)\right\rangle/\left\langle I(t)\right\rangle^2$ vanishes at
$\tau=0$~ns. Figure 2(a) shows the measured $g^{(2)}(\tau)$ function of a
single nanodiamond. The interval between the peaks represents the repetition time of the
pulsed laser. In contrast to what is expected from classical light there is no
peak at $\tau = 0$~ns, showing that the diamond contains only a single
NV-center. Using our recently developed pick-and-place technique
\cite{Schroeder2010} this pre-selected diamond was picked up with an atomic
force microscope (AFM) and preliminary placed on the GaP surface, close to the
PCCs. Here the diamond was further characterized. In particular, a spectrum was
measured [Fig.~2(b)] and the position of the ZPL at $639.5$ nm was determined
[red dots in Fig.~2(c)].

We selected a cavity with a resonance wavelength of $642.9$~nm and a $Q$-factor
of $Q = 1003$ and performed active tuning to the ZPL of this diamond in the
following way: A focused blue laser (407~nm, 270~$\mu$W), which is absorbed
within the GaP membrane, was used to heat the structure locally for several
minutes. It is assumed that by this procedure the GaP locally oxidizes and
changes its index of refraction.\cite{Schmidt2009} After this procedure the
cavity's resonance was blue shifted by $3.4$~nm and showed a resonance at
$639.5$~nm. The quality factor decreased from $Q = 1003$ to $Q = 603$.
Figure~2(c) shows the fluorescence spectra of the bare cavity after tuning and
the fluorescence from the NV-center. Note that the cavity resonance wavelength
matches the ZPL almost perfectly.

In a last step the diamond was picked up again using the AFM and placed now in
the center of the cavity [Fig.~1(a)]. Since the field strength of the cavity
mode decays exponentially outside the slab, it is crucial to place the
NV-center as close as possible to the surface.\cite{Lukin2010} In the AFM
images the diamond shows a height of about 35~nm, meaning that the NV-center is
less than 35 nm above the surface. Figure~3 shows the measured spectra of the
cavity with and without the nanodiamond, both taken at 300~$\mu$W excitation
power. The pronounced enhancement of the ZPL emission at 639.5 nm is striking.
The autocorrelation measurement (inset in Fig.~3) proves that the light emitted
from the cavity has a strong non-classical behaviour, revealing that the
fluorescence mainly originates from the NV-center. The peaks around 610~nm stem
from higher-order modes with smaller $Q$-factors. As no influence of the
NV-center on the fluorescence intensity is expected below 620~nm, the two
spectra were normalized with respect to the emission at 600~nm to compensate
for a slightly different alignment. This allows to reveal which part of the
cavity's fluorescence (red curve in Fig.~3) has it's origin in the NV-center by
subtracting the (normalized) bare cavity background (green curve in Fig.~3).

To quantify the observed enhancement in the ZPL, we calculate the spectrally
resolved SE enhancement due to the cavity $F^*$ by comparing the ZPL emission
in the cavity and on the bare substrate. We would like to point out that care
has to be taken when comparing the optical properties of a bare defect center
and one coupled to a photonic structure. Any change in the dielectric
environment also influences the corresponding emission properties such as the
overall SE rate \cite{Lukin2010}. However, the spectral shape is changed only
near the cavity resonances. This allows us to normalize the measured NV-spectra [Fig. 2(c) and Fig. 3] to
the broad spectral region above and below the ZPL. Afterwards, a Lorentzian was
fitted to the data, resulting in (normalized) peak intensities of
$I^*_{sub}=1.2$ and $I^*_{cav}=14.9$ on the substrate and on the cavity,
respectively. This yields the experimental Purcell enhancement $F^* =
I^{*}_{cav}/I^{*}_{sub} = 12.1$. The difference to the theoretical value $F =
61$, calculated from Eq.~(1), is caused by non-ideal alignment. Neither is the
NV-center placed in the field maximum of the cavity mode, nor is the dipole
moment's orientation optimized.

A further enhancement of the emission into the ZPL can be achieved by improving
the $Q$-factor or by performing experiments at cryogenic temperatures. At
4~Kelvin the ZPL can be nearly Fourier-limited \cite{Wang2008} and about 3$\%$
of the light is emitted into the ZPL. In this case we estimate that coupling to
a similar PCC with a $Q$-factor of 600 should allow channeling of almost $30\%$
of the emission into the cavity mode.

In conclusion, we have demonstrated the deterministic coupling of the zero phonon
line of a single nitrogen-vacancy center in a nanodiamond to a photonic crystal
cavity. This is a major step towards the realization of integrated quantum
optical devices. With the presented pick-and-place technique and the selective
cavity tuning, even more complex systems, involving two cavities and emitters,
\cite{Hollenberg2008b} can be assembled in a controlled way. Simple quantum
gates, integrated on a single photonic crystal chip, are within reach.

This work was supported by the DFG (BE2224/9) and the BMBF (KEPHOSI).
J.~Wolters acknowledges funding by the state of Berlin (NaF\"{o}G).

\clearpage

\begin{figure}[h!]
\centering
\includegraphics[width=0.4\textwidth]{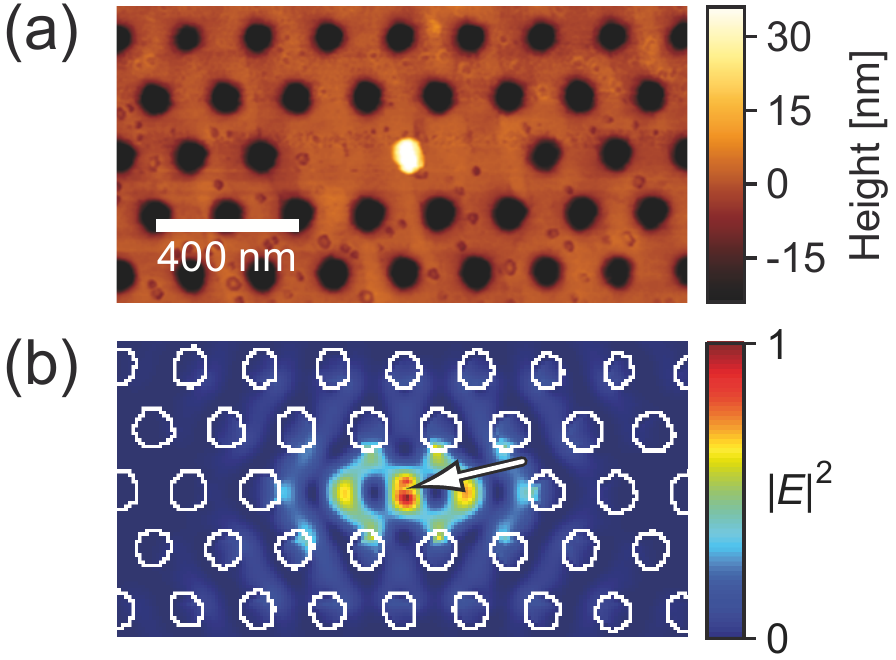}
 \caption{(Color online)
(a) AFM image of the L3 GaP cavity with a nanodiamond of 35 nm height located close to the center.
The lattice constant is 200 nm.
(b) Simulated (FDTD) electric field profile of the fundamental mode of this cavity. The geometry was adapted from the AFM image in (a). The arrow indicates the position of the diamond.}
 \label{fig:AFM}
\end{figure}
\clearpage
\begin{figure}[h]
\centering
\includegraphics[width=0.4\textwidth]{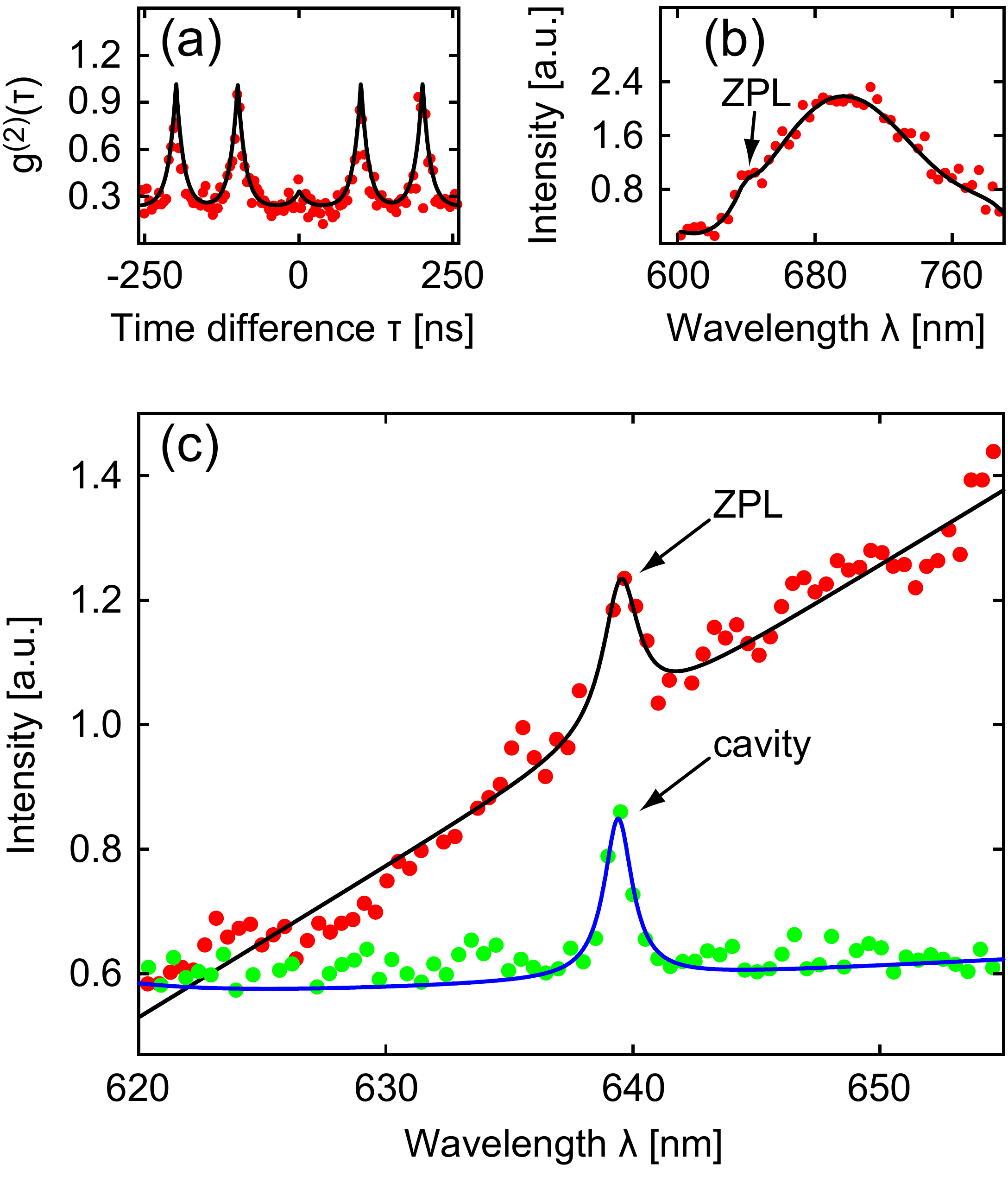}
\caption{(Color online)
(a) Autocorrelation measurement from a nanodiamond on a glass cover slip under pulsed excitation, clearly revealing the single photon character of the fluorescence. The black curve is a fit to the data.
(b) Full spectrum of the emission from a single NV-center in a nanodiamond on the GaP substrate under cw-excitation. The black curve is a guide to the eye.
(c) Red dots: Zoom into the ZPL of the same NV-center as in (b). The black curve is a fit to the data.
Green dots: Corresponding measurement of the cavity fluorescence spectrum.
The blue line is a fit to the data, showing almost perfect matching with the spectral position of the ZPL.
For better visibility an offset of $0.5$ is added to the data.
}
\label{fig:Antibunching}
\end{figure}

\clearpage
\begin{figure}[h]
\centering
\includegraphics[width=0.4\textwidth]{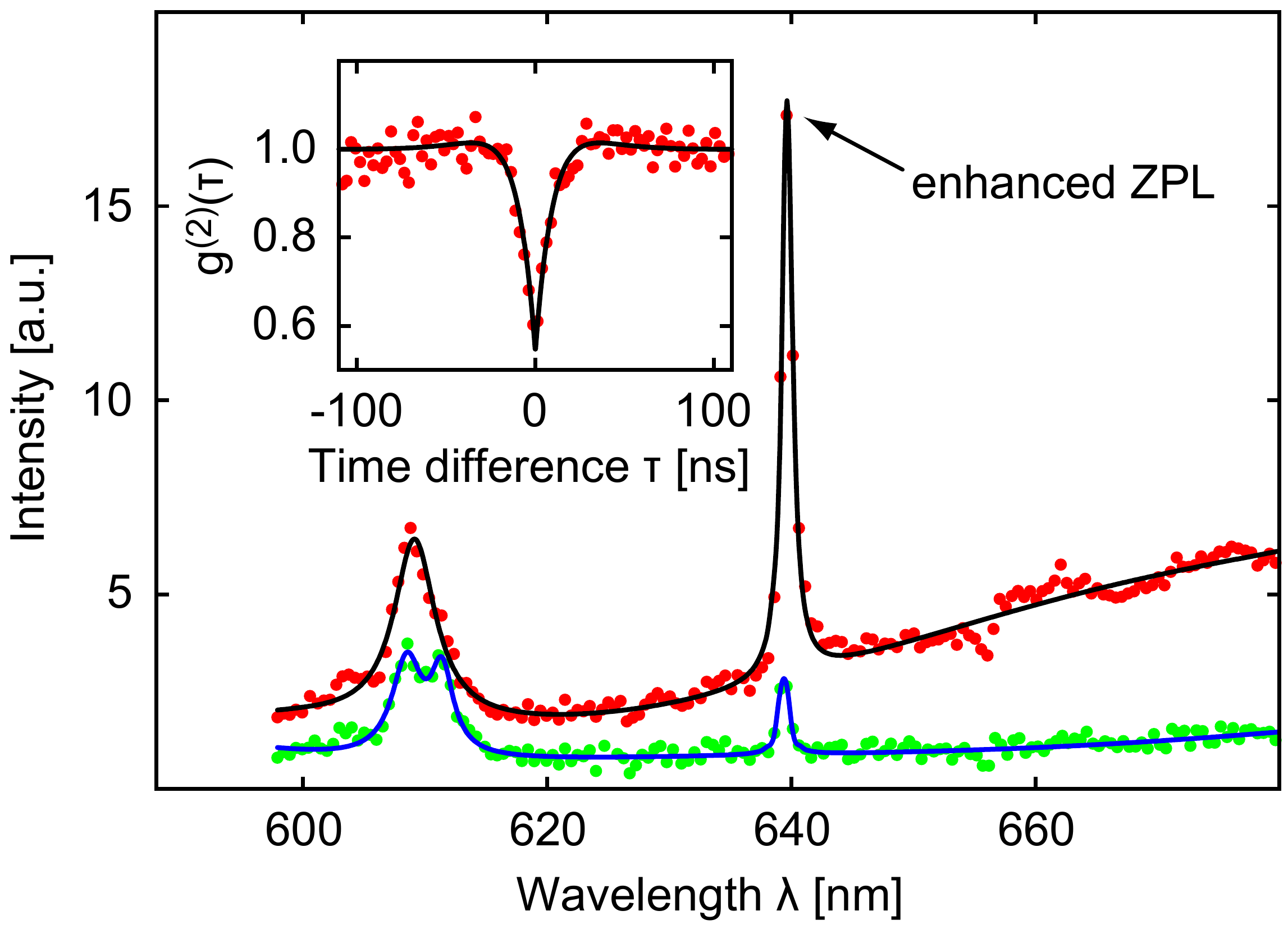}
\caption{(Color online)
Green dots: Normalized spectrum of the cavity fluorescence without nanodiamond.
The peaks around 610 nm stem from higher-order modes.
The blue curve is a fit to the data.
Red dots: Normalized fluorescence spectrum with nanodiamond.
For better visibility an offset of 1 is added.
The black curve is a fit to the data.
Inset: Autocorrelation measurement of the cavity with nanodiamond.
The dip at $\tau = 0$~ns proves the non-classical behaviour of the light emitted from the assembled system.
}
 \label{fig:}
\end{figure}

\end{document}